\documentclass[twocolumn,showpacs,preprintnumbers,amsmath,amssymb]{revtex4}

\usepackage{amsfonts}

\usepackage{graphicx}
\usepackage{dcolumn}
\usepackage{bm}

\begin{document}

\title{Quantum information approach to the quantum phase transition in the Kitaev honeycomb model}

\author{Jian Cui }
\email{cuijian@iphy.ac.cn}
\author{Jun-Peng Cao}%
\author{Heng Fan}
\affiliation{%
Institute of Physics, Chinese Academy of Sciences, Beijing National
Laboratory for Condensed Matter Physics, Beijing 100190, China
}%

\date{\today}

\begin{abstract}

Kitaev honeycomb model with topological phase transition at zero
temperature is studied using quantum information method. Based on
the exact solution of the ground state, the mutual information
between two nearest sites and between two bonds with longest
distance are obtained. It is found that the mutual information shows
some singularities at the critical point the system transits from
gapless phase to gapped phase. The finite-size effects and scaling
behavior are also studied. Our results indicate that the mutual
information can serve as good indicator of the topological phase
transition. This is because that the mutual information is believed
to be able to catch some global correlation properties of the
system. Meanwhile, this method has advantages that the phase
transition can be determined easily and the order parameters, which
are hard to be obtained for some topological phase transitions, are
not necessarily known.

\end{abstract}


\maketitle

\section{\label{sec:level1} I\lowercase{ntroduction}}

Recently, Kitaev honeycomb model has become a popular subject in
both the fields of condensed matter physics and quantum information
processing
\cite{gkellsjvala,DHLXT,smdsks,kpssdjv,sdkpsjv,jvkpssd,gbsmrs,dfapzanadi,xyfxt,gkjv,smns,jhzhqz,xfsfn}.
This model was first introduced by Kitaev to study the anyons, and
the analytic exact solution to the ground state of this model has
been obtained by several
methods\cite{kitaev,xyfxt,shuoyang,zngo,hdczn}. It has rich phase
transitions and has both a gapless phase with non-Abelian anyons
excitation and three gapped phases with Abelian anyons excitations
depending on the values of the parameters in the Hamiltonian. It is
shown that the system possess a topological phase transition, which
is not able to be characterized by symmetry breaking theory and the
corresponding local order parameters but can be characterized by
nonlocal string order parameters\cite{xyfxt,xgwen}. One interesting
point is that the system is a scarce exactly solvable model with
dimensions higher than one , thus it provides a test bed for many
numerical methods in two dimensional systems just as the Ising model
does in one dimension. With these interesting properties, Kitaev
honeycomb model has been studied intensively and is extended to
other cases\cite{hysak,smns}.

On the other hand, Kitaev honeycomb model also has good practical
advantages to be an active subject in that it has great potential
applications in the quantum information and quantum computation. It
was suggested to use the Kitaev honeycomb model to realize the
fault-tolerant topological quantum computation. The system is a good
candidate to encode quantum information while those quantum states
can be naturally protected from the inevitable decoherence by
environment\cite{kitaev03}.
  The Kitaev
honeycomb model can be realized using the optical
lattice\cite{amgkbpz,duan}, and using the superconducting quantum
circuits\cite{zyx,jqy}. It is also studied by means of fidelity
susceptibility\cite{shuoyang} and the extended Kitaev model is
studied by approaches of entanglement\cite{xfsfn}.

In this paper, we investigate the Kitaev honeycomb model from the
quantum information perspective\cite{nielsen,vedral}. We study the
topological phase transition in this model by means of mutual
information between the component lattices. It is generally believed
that the mutual information measures the total information and
describes the global correlation properties\cite{winter}. We find
that both the derivative of mutual information between two nearest
neighbor lattices and the mutual information between two bonds of
the lattice can detect the topological phase transition in the
Kitaev honeycomb model. This quantum information method has great
advantages in that the singular behavior occurs exactly at the point
when the gapless phase transits into a gapped phase. We also study
the finite-size effects and the scaling behavior of the
singularities of the mutual information.

This paper is organized as follows. In Section II we briefly
introduce the Kitaev honeycomb model, then diagonalize the
Hamiltonian and give the exact solution of the ground state based on
the initial Kitaev's method. After that, we calculate the two sites
and four sites correlation functions getting prepared for the
two-site and two-bond reduced density matrix. In section III and
section IV, we calculate the two-site mutual information and
two-bond mutual information and the former one's derivative,
respectively. Section V is the conclusions and remarks.

\section{\label{sec:level1}K\lowercase{itaev honeycomb model } }
Kitaev honeycomb model is a two dimensional spin-$\frac{1}{2}$
lattice model with nearest neighbor interactions. It has two kinds
of simple sublattice which are denoted by the dark dots and empty
circles in figure 1. Each lattice interacts with three nearest
neighbors of the opposite kind through three distinct bonds labeled
as $x$ link, $y$ link and $z$ link. For each bond the interaction
has different coupling constant. The Hamiltonian is
\begin{equation}H=-J_x\sum_{x-links}\sigma^x_j\sigma^x_k-J_y\sum_{y-links}\sigma^y_j\sigma^y_k-J_z\sum_{z-links}\sigma^z_j\sigma^z_k,
\end{equation}
where the subindex $j,k$ denote the location of the site, and
$\sigma^\alpha_k(\alpha=x,y,z)$ is the pauli matrix at site $k$. We
take axis of the system in the $\textbf{n}_1$ and $\textbf{n}_2$
direction, and in each direction there are $L$ unit cells. Therefore
the whole system has $2L^2$ sites. Next, we used the original
Kitaev's method to diagonalize this Hamiltonian and get its ground
state.

\begin{figure}
\includegraphics[height=6.8cm,width=7cm]{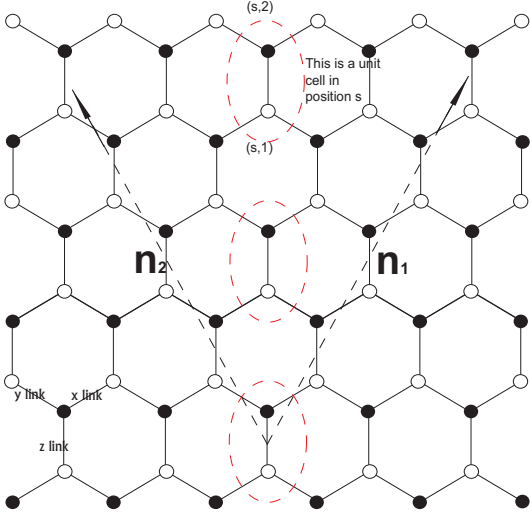}
\caption{\label{fig:epsart}(color online). The sketch map of Kitaev
honeycomb model. The unit cell contains two sites of different
kinds, which is highlighted by a elliptic circle. For simplicity, we
choose the coordinate axes in $\textbf{n}_1$ and $\textbf{n}_2$
directions.}
\end{figure}

\subsection{The ground state}
We first introduce the following Majorana transformation to
transform the Pauli operators into the Majorana fermion operators.
\begin{equation}
\sigma^x=ib^xc,  \sigma^y=ib^yc,  \sigma^z=ib^zc,
\end{equation}
where the Majorana operators satisfy $A^\dag=A$, $A^2=1$, $AB+BA=0$,
$b^xb^yb^zc=1$, for $A,B\in \{b^x,b^y,b^z,c\}$ and $A\neq B$. Thus,
the Hamiltonian becomes
\begin{eqnarray}
H&=&-\sum_{\alpha}J^{\alpha}\sum_{\alpha-links}b_j^{\alpha}b_k^{\alpha}c_jc_k\nonumber\\
&=&i\sum_{\alpha}J^{\alpha}\sum_{\alpha-links}(ib_j^{\alpha}b_k^{\alpha})c_jc_k\nonumber\\
&=&i\sum_{\alpha}J^{\alpha}\sum_{\alpha-links}\hat{u}_{jk}c_jc_k\nonumber\\
&=&\frac{i}{2}\sum_{j,k}J_{{\alpha}_{j,k}}\hat{u}_{j,k}c_jc_k.
\end{eqnarray}
In the last equation, the value of $\alpha$ is totally determined by
the site index $j$ and $k$. The factor $\frac{1}{2}$ is due to the
summation of the lattices has counted each lattice twice. It can be
easily shown that $\hat{u}_{j,k}^2=1,[\hat{u}_{j,k},H]=0,$ and
$\hat{u}_{j,k}$ commute with each other. As a result the eigenvalues
of $\hat{u}_{j,k}$ here we present by $u_{j,k}$ are $\pm1$, and the
whole Hilbert space can be decomposed into a series of eigenvalue
spaces described by the eigenvalues of $\hat{u}_{j,k}$. According to
\cite{libe,kitaev}, the ground state is in the vortex free space so
that we assume $u_{j,k}=1$ for all links, where $j$ is a kind of
simple sublattice presented by the empty circles in this paper.
Notice $u_{j,k}=-u_{k,j}$.

As the unit cell of this model contains one empty circle lattice and
one dark dot lattice, we introduce a pair of index
$(\textbf{s},\lambda)$ to take the place of the previous site index
$j$, where the first index $\textbf{s}$ stands for the location of
the unit cell, and the second one describes the two different kinds
of sublattice. In this paper, we let the empty circle's second index
takes the value $1$, and the dark dot's takes the value $2$. Please
see figure 1. Then the Hamiltonian becomes
\begin{eqnarray}
 H&=&\frac{i}{2}\sum_{\textbf{s},\lambda,\textbf{t},\mu}J_{\textbf{s},\lambda,\textbf{t},\mu}c_{\textbf{s},\lambda}c_{\textbf{t},\mu}.
\end{eqnarray}
The two dimensional system we studied is on the surface of a torus
with periodical boundary conditions. Because of the translational
invariance of the system $J_{\textbf{s},\lambda,t,\mu}$ is actually
determined by three index $\lambda,\mu,$ and
$\textbf{t}-\textbf{s}$. Then we introduce the Fourier
transformation
\begin{eqnarray}
J_{\textbf{s},\lambda;\textbf{t},\mu}&=&J_{0,\lambda;\textbf{t}-\textbf{s},\mu}
=\frac{1}{L^2}\sum_{\textbf{q}}e^{-i\textbf{q}\cdot(\textbf{r}_t-\textbf{r}_s)}\widetilde{J_{\lambda,\mu}}(\textbf{q}),\nonumber\\
c_{\textbf{s},\lambda}&=&\sqrt{\frac{2}{L^2}}\sum_{\textbf{q}}e^{i\textbf{q}\cdot
\textbf{r}_s}a_{\textbf{q},\lambda}.
\end{eqnarray}
The inverse transformation is
\begin{eqnarray}
\widetilde{J_{\lambda,\mu}}(\textbf{q})&=&\sum_\textbf{t}e^{i\textbf{q}\cdot
\textbf{r}_t}J_{0,\lambda;\textbf{t},\mu},\nonumber\\
a_{{\textbf{q}},\lambda}&=&\sqrt{\frac{1}{2L^2}}\sum_\textbf{s}e^{-i\textbf{q}\cdot\textbf{r}_s}c_{\textbf{s},\lambda},
\end{eqnarray}
where $a_{\textbf{q},\lambda}$ satisfies
$a_{-\textbf{q},\lambda}=a_{\textbf{q},\lambda}^{\dag}$,
$a_{\textbf{q},\lambda}^2=0$,
$[a_{\textbf{q},\lambda},a_{\textbf{q},\mu}^{\dag}]_{+}\equiv
a_{\textbf{q},\lambda}a_{\textbf{q},\mu}^{\dag}+a_{\textbf{q},\mu}^{\dag}a_{\textbf{q},\lambda}=\delta_{\textbf{p}\textbf{q}}\delta_{\lambda,\mu}
$, and other anticommutators are all equal to zero. Then the
Hamiltonian arrives at
\begin{eqnarray}
H&=&i\sum_{\textbf{q}}\sum_{\lambda,\mu=1}^2\widetilde{J_{\lambda,\mu}}(\textbf{q})a_{-\textbf{q},\lambda}a_{\textbf{q},\mu}.
\end{eqnarray}
After simple calculations we obtain that
$\widetilde{J_{1,1}}(\textbf{q})=\sum_te^{i\textbf{q}\cdot
\textbf{r}_t}J_{01,\textbf{t}1}=0$, because $J_{01,\textbf{t}1}=0$.
For the similar reason $\widetilde{J_{2,2}}(\textbf{q})=0$. As each
lattice interacts with its three nearest neighbors, there are only
three values of $\textbf{t}$ corresponding to the three neighbors
that make $J_{01,\textbf{t}2}$ take nonzero values. Thus
$\widetilde{J_{1,2}}(\textbf{q})=J_xe^{i\textbf{q}\cdot\textbf{n}_1}+J_ye^{i\textbf{q}\cdot\textbf{n}_2}+J_z$,
and
$\widetilde{J_{2,1}}(\textbf{q})=-\widetilde{J_{1,2}}^*(\textbf{q})$,
where $\textbf{n}_1$ and $\textbf{n}_2$ are in certain directions
which is shown in figure 1. Let
$f(\textbf{q})\equiv\widetilde{J_{1,2}}(\textbf{q})=\varepsilon(\textbf{q})+i\Delta(\textbf{q})$,
and choose $\overrightarrow{q_x}$ to be in the direction of
$\textbf{n}_1$, and $\overrightarrow{q_y}$ to be in the direction of
$\textbf{n}_2$. Then we have
\begin{eqnarray}
\varepsilon(\textbf{q})&=&J_xcosq_x+J_ycosq_y+J_z\nonumber,\\
\Delta(\textbf{q})&=&J_xsinq_x+J_ysinq_y ,
\end{eqnarray}
where $q_x$ and $q_y$ take values $q_x,q_y=2\pi n/L$,
$n=-(L-1)/2,\cdots,(L-1)/2$.
 We can see that
$\varepsilon(-\textbf{q})=\varepsilon(\textbf{q})$,
$\Delta(-\textbf{q})=-\Delta(\textbf{q})$, and
$f(-\textbf{q})=f^*(\textbf{q})$. The Hamiltonian becomes
\begin{eqnarray}
H=\sum_{\textbf{q}}if(\textbf{q})a^{\dag}_{\textbf{q},1}a_{\textbf{q},2}+(if(\textbf{q}))^*a^{\dag}_{\textbf{q},2}a_{\textbf{q},1}.
\end{eqnarray}
Next, we introduce the following Bogoliubov transformation:
\begin{eqnarray}
C_{\textbf{q},1}&=&u_{\textbf{q}}a_{\textbf{q},1}+v_{\textbf{q}}a_{\textbf{q},2}\nonumber,\\
C_{\textbf{q},1}^{\dag}&=&u_{\textbf{q}}^{*}a_{\textbf{q},1}^{\dag}+v_{\textbf{q}}^{*}a_{\textbf{q},2}^{\dag}\nonumber,\\
C_{\textbf{q},2}&=&v_{\textbf{q}}^{*}a_{\textbf{q},1}-u_{\textbf{q}}^{*}a_{\textbf{q},2}\nonumber,\\
C_{\textbf{q},2}^{\dag}&=&v_{\textbf{q}}a^{\dag}_{\textbf{q},1}-u_{\textbf{q}}a^{\dag}_{\textbf{q},2},
\end{eqnarray}
with the new operators satisfying
$[C_{\textbf{q},\lambda},C_{\textbf{p},\mu}^{\dag}]_{+}=\delta_{\textbf{p}\textbf{q}}\delta_{\lambda,\mu}$,$C_{\textbf{q},\lambda}^2=0$.
Using the Bogoliubov transformation, the Hamiltonian is diagonalized
as
\begin{eqnarray}
H=\sum_{\textbf{q}}|f_{\textbf{q}}|(C_{\textbf{q},1}^{\dag}C_{\textbf{q},1}-C_{\textbf{q},2}^{\dag}C_{\textbf{q},2}),
\end{eqnarray}
with $u_\textbf{q}=\frac{1}{\sqrt{2}}$,
$v_{\textbf{q}}=\frac{i}{\sqrt{2}}\frac{f_{\textbf{q}}}{|f_{\textbf{q}}|}$,
$v_{-\textbf{q}}=-v_{\textbf{q}}^*$,
$C_{-\textbf{q},1}=-2u_{\textbf{q}}^*v_{\textbf{q}}^*C_{\textbf{q},2}^{\dag}$.
Considering the fact
$C_{\textbf{q},1}^{\dag}C_{\textbf{q},1}=1-C_{-\textbf{q},2}^{\dag}C_{-\textbf{q},2}$,
the Hamiltonian reads
\begin{eqnarray}
H&=&\sum_{\textbf{q}}|f_{\textbf{q}}|(1-C_{-\textbf{q},2}^{\dag}C_{-\textbf{q},2}-C_{\textbf{q},2}^{\dag}C_{\textbf{q},2})\nonumber\\
&=&\sum_{\textbf{q}}|f_{\textbf{q}}|(1-2C_{\textbf{q},2}^{\dag}C_{\textbf{q},2}).
\end{eqnarray}

The normalized ground state is
\begin{eqnarray}
|G\rangle=\prod_{\textbf{q}}C_{\textbf{q},2}^{\dag}|0\rangle,
\end{eqnarray}
 with $C_{\textbf{q},2}|0\rangle=0$.
The energy gap is $2\min_{\textbf{q}}\{|f_{\textbf{q}}|\}$.

\subsection{Phase diagram }

This ground state has two distinct phases in the parameter space. In
the region of $ |J_x|\leq|J_y|+|J_z|$, $ |J_y|\leq|J_x|+|J_z|$ and $
|J_z|\leq|J_y|+|J_x|$ it is gapless with non-Abelian excitation and
in other regions it is gapped with Abelian anyon
excitations\cite{kitaev}. We focus on the $J_x+J_y+J_z=1$ plane. The
phase diagram is shown in figure $2$. In this paper, we investigate
the behaviors of two-site mutual information and two-bond mutual
information in the phase transition from the gapless phase to a
gapped phase along the red dash line in the phase diagram of figure
$2$.

\begin{figure}
\includegraphics[height=7cm,width=7cm]{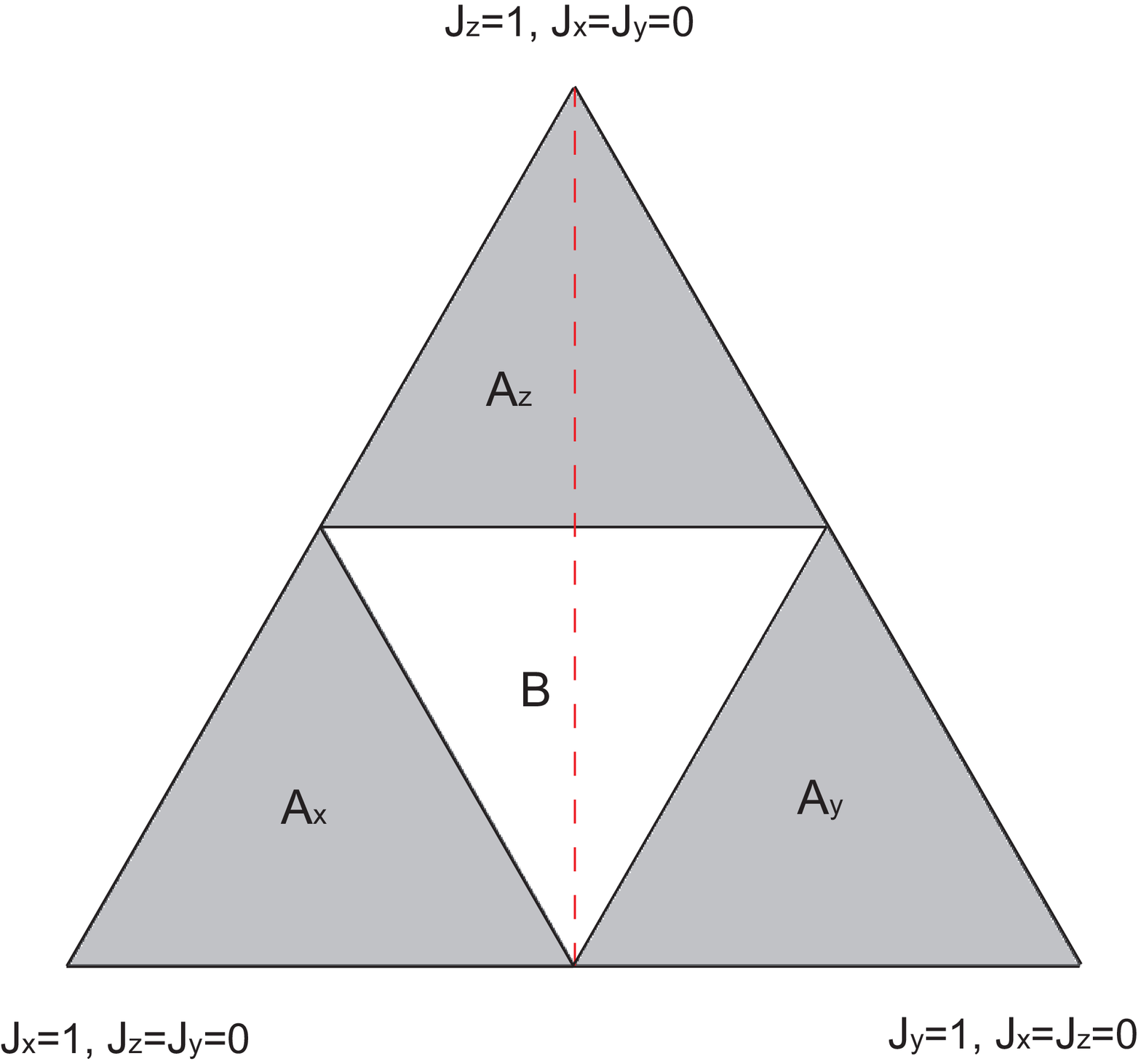}
\caption{\label{fig:epsart}(color online). The phase diagram of the
Kitaev honeycomb model in the $J_x+J_y+J_z=1$ plane in the parameter
space. In the three shadow areas labeled by $A_x$, $A_y$ and $A_z$,
the system is gapped with Abelian anyon excitation, and in the blank
area labeled by $B$ the system is gapless with non-Abelian
excitation. In this paper, we focus on the red dash line
$J_x=J_y=(1-J_z)/2$, where the critical point of topological phase
transition is $J_z=0.5$.}
\end{figure}

\subsection{Correlation functions}
In this section we calculate the two-site and four-site correlation
functions at the ground state of the systems which will be used to
construct the reduced density matrix. Suppose the two nearest
lattices to be studied are linked by $z$-bonds. The correlation
function between two nearest lattices is
\begin{eqnarray}
\langle\sigma^z_{\textbf{r},1}\sigma^z_{\textbf{r},2}\rangle&=&\langle b^z_{\textbf{r},1}b^z_{\textbf{r},2}\frac{2}{L^2}\sum_{\textbf{q},\textbf{q}^{'}}e^{i(\textbf{q}+\textbf{q}^{'})\cdot\textbf{r}}a_{\textbf{q},1}a_{\textbf{q}^{'},2}\rangle\nonumber\\
&=&-i\frac{2}{L^2}\sum_{\textbf{q},\textbf{q}^{'}}e^{i(\textbf{q}+\textbf{q}^{'})\cdot\textbf{r}}\langle
a_{\textbf{q},1}a_{\textbf{q}^{'},2}\rangle\nonumber.
\end{eqnarray}
By using the relation
\begin{eqnarray}
\langle a_{\textbf{q},1}a_{\textbf{q}^{'},2}\rangle&=&\langle(u^*_{\textbf{q}}C_{\textbf{q},1}+v_{\textbf{q}}C_{\textbf{q},2})(v^*_{\textbf{q}^{'}}C_{\textbf{q}^{'},1}-u_{\textbf{q}^{'}}C_{\textbf{q}^{'},2})\rangle\nonumber\\
&=&-u^*_{\textbf{q}}u_{\textbf{q}^{'}}\langle C_{\textbf{q},1}C_{\textbf{q}^{'},2}\rangle\nonumber\\
&=&\frac{i}{2}\delta_{\textbf{q},-\textbf{q}^{'}}\frac{f_{\textbf{q}}}{|f_{\textbf{q}}|}\nonumber,
\end{eqnarray}
we obtain the correlation function
\begin{eqnarray}
\langle\sigma^z_{\textbf{r},1}\sigma^z_{\textbf{r},2}\rangle&=&\frac{1}{L^2}\sum_{{\textbf{q}}}\frac{f_{\textbf{q}}}{|f_{\textbf{q}}|}=\frac{1}{2L^2}\sum_{{\textbf{q}}}\frac{f_{\textbf{q}}+f_{-\textbf{q}}}{|f_{\textbf{q}}|}\nonumber\\
&=&\frac{1}{L^2}\sum_{\textbf{q}}\frac{\varepsilon_{\textbf{q}}}{E_{\textbf{q}}},
\end{eqnarray}
where
$E_{\textbf{q}}=|f_{\textbf{q}}|=\sqrt{\varepsilon^2_{\textbf{q}}+\Delta^2_{\textbf{q}}}$.

The correlation function between two bonds as highlighted by
elliptic circles in figure 1 is
\begin{widetext}
\begin{eqnarray}
\langle\sigma^z_{\textbf{r}_1,1}\sigma^z_{\textbf{r}_1,2}\sigma^z_{\textbf{r}_2,1}\sigma^z_{\textbf{r}_2,2}\rangle&=&\langle b^z_{\textbf{r}_1,1}b^z_{\textbf{r}_1,2}b^z_{\textbf{r}_2,1}b^z_{\textbf{r}_2,2}C_{\textbf{r}_1,1}C_{\textbf{r}_1,2}C_{\textbf{r}_2,1}C_{\textbf{r}_2,2}\rangle\nonumber\\
&=&-\frac{4}{L^4}\sum_{\textbf{q}_1,\textbf{q}_2,\textbf{q}_3,\textbf{q}_4}e^{i(\textbf{q}_1+\textbf{q}_2)\cdot
\textbf{r}_1}e^{i(\textbf{q}_3+\textbf{q}_4)\cdot
\textbf{r}_2}\langle
a_{\textbf{r}_1,1}a_{\textbf{r}_1,2}a_{\textbf{r}_2,1}a_{\textbf{r}_2,2}\rangle\nonumber,
\end{eqnarray}
 where
\begin{eqnarray}
\langle a_{\textbf{r}_1,1}a_{\textbf{r}_1,2}a_{\textbf{r}_2,1}a_{\textbf{r}_2,2}\rangle&=&-\frac{1}{4}\frac{f_{\textbf{q}_1}}{|f_{\textbf{q}_1}|}\frac{f_{\textbf{q}_3}}{|f_{\textbf{q}_3}|}\langle C_{-\textbf{q}_1,2}^{\dag}(C_{-\textbf{q}_2,2}+C_{\textbf{q}_2,2})(C^{\dag}_{-\textbf{q}_3,2}-C_{\textbf{q}_3,2})C_{\textbf{q}_4,2}\rangle\nonumber\\
&=&\frac{1}{4}\frac{f_{\textbf{q}_1}}{|f_{\textbf{q}_1}|}\frac{f_{\textbf{q}_3}}{|f_{\textbf{q}_3}|}(\delta_{\textbf{q}_2,-\textbf{q}_3}\delta_{\textbf{q}_1,-\textbf{q}_4}-\delta_{\textbf{q}_1,-\textbf{q}_3}\delta_{\textbf{q}_2,-\textbf{q}_4}-\delta_{\textbf{q}_1,-\textbf{q}_2}\delta_{\textbf{q}_3,-\textbf{q}_4})\nonumber.
\end{eqnarray}
Then, we arrive at
\begin{eqnarray}
\langle\sigma^z_{\textbf{r}_1,1}\sigma^z_{\textbf{r}_1,2}\sigma^z_{\textbf{r}_2,1}\sigma^z_{\textbf{r}_2,2}\rangle&=&-\frac{1}{L^4}\big(\sum_{\textbf{q}_1,\textbf{q}_3}\frac{f_{\textbf{q}_1}}{|f_{\textbf{q}_1}|}\frac{f_{\textbf{q}_3}}{|f_{\textbf{q}_3}|}e^{i(\textbf{q}_1-\textbf{q}_3)\cdot(\textbf{r}_1-\textbf{r}_2)}-\sum_{\textbf{q}_1,\textbf{q}_2}e^{i(\textbf{q}_1+\textbf{q}_2)\cdot(\textbf{r}_1-\textbf{r}_2)}-\sum_{\textbf{q}_1,\textbf{q}_3}\frac{f_{\textbf{q}_1}}{|f_{\textbf{q}_1}|}\frac{f_{\textbf{q}_3}}{|f_{\textbf{q}_3}|}\big)\nonumber\\
&=&-\frac{1}{L^4}\sum_{\textbf{q}_1,\textbf{q}_3}\frac{f_{\textbf{q}_1}f_{\textbf{q}_3}+f_{-\textbf{q}_1}f_{-\textbf{q}_3}}{|f_{\textbf{q}_1}|\cdot|f_{\textbf{q}_3}|}\big(\cos[(\textbf{q}_1-\textbf{q}_3)\cdot(\textbf{r}_1-\textbf{r}_2)]-1\big)\nonumber\\
&=&\frac{1}{L^4}\sum_{\textbf{q}_1,\textbf{q}_3}\frac{\Delta_{\textbf{q}_1}\Delta_{\textbf{q}_3}-\varepsilon_{\textbf{q}_1}\varepsilon_{\textbf{q}_3}}{E_{\textbf{q}_1}E_{\textbf{q}_3}}\big(\cos[(\textbf{q}_1-\textbf{q}_3)\cdot(\textbf{r}_1-\textbf{r}_2)]-1\big).
\end{eqnarray}

\end{widetext}

\section{\label{sec:level1}M\lowercase{utual information between two neighbor lattices}   }

\begin{figure}
\includegraphics[height=5.3cm,width=\linewidth]{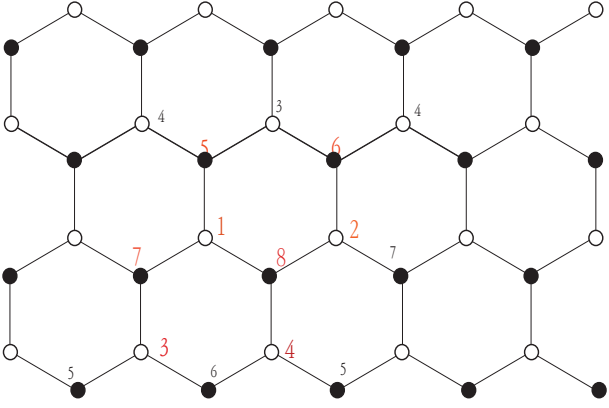}
\caption{\label{fig:epsart}(color online). The smallest translation
invariant subsystem of Kitaev honeycomb model with periodical
boundary conditions. The eight interaction sites in the subsystem
are highlighted by red numbers in the graph. The sites labeled by
small black numbers are the repetitions of the $8$ sites because of
the periodical boundary condition and torus topology. }
\end{figure}
\begin{figure}
\includegraphics[height=7cm,width=\linewidth]{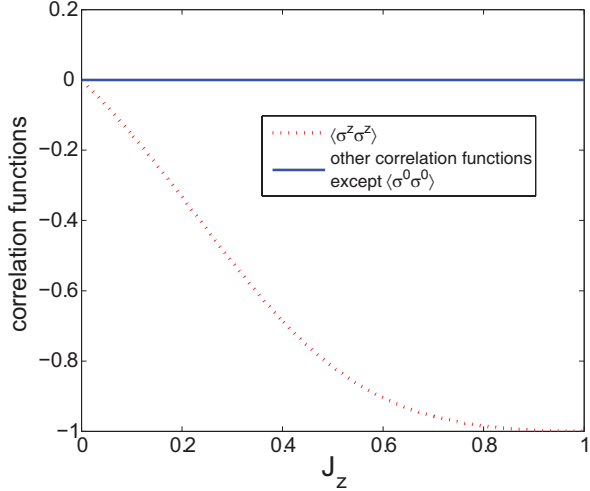}
\caption{\label{fig:epsart}(color online). The ground state
correlation functions of two-site with $z$-link in the Kitaev
honeycomb model. We see that
$\langle\sigma_{\textbf{r},1}^{0}\sigma_{\textbf{r},2}^{0}\rangle=1$,
$\langle\sigma_{\textbf{r},1}^{z}\sigma_{\textbf{r},2}^{z}\rangle
\neq 0$ and others are zero. $J_z$ is in unit of $J_x+J_y+J_z$.}
\end{figure}

The reduced density matrix of two site $i$ and $j$ is
$\rho_{i,j}=\frac{1}{4}\sum_{\alpha,\beta=0}^3\langle\sigma_i^{\alpha}\sigma_j^{\beta}\rangle\sigma_i^{\alpha}\sigma_j^{\beta}$,
where $\sigma^1=\sigma^x$, $\sigma^2=\sigma^y$, $\sigma^3=\sigma^z$
and $\sigma^0$ is identity. In the system (1), each site interacts
with its three neighbors by different operators($\sigma^x$,
$\sigma^y$ and $\sigma^z$), while each two linked sites together
have only one kind of operator available, i.e.,
$\sigma^{\alpha}\sigma^{\alpha}$, with $\alpha$ corresponding to the
type of their link. We find that only the correlation function along
the link interacting direction is nonzero when we study the reduced
matrix of two linked sites. That is to say if we consider the two
lattices with $z$ link (please see figure 1), all the correlations
are zero except $\langle\sigma^z\sigma^z\rangle$. Therefore, the
reduced density matrix of this model has only diagonal elements,
although the interactions of one site have three components. It
indicates that the model (1) is much more like a classical system
and similar to the Ising model. That may explain why this two
dimensional model can be solved analytically.

\begin{figure}
\includegraphics[height=7cm,width=8cm]{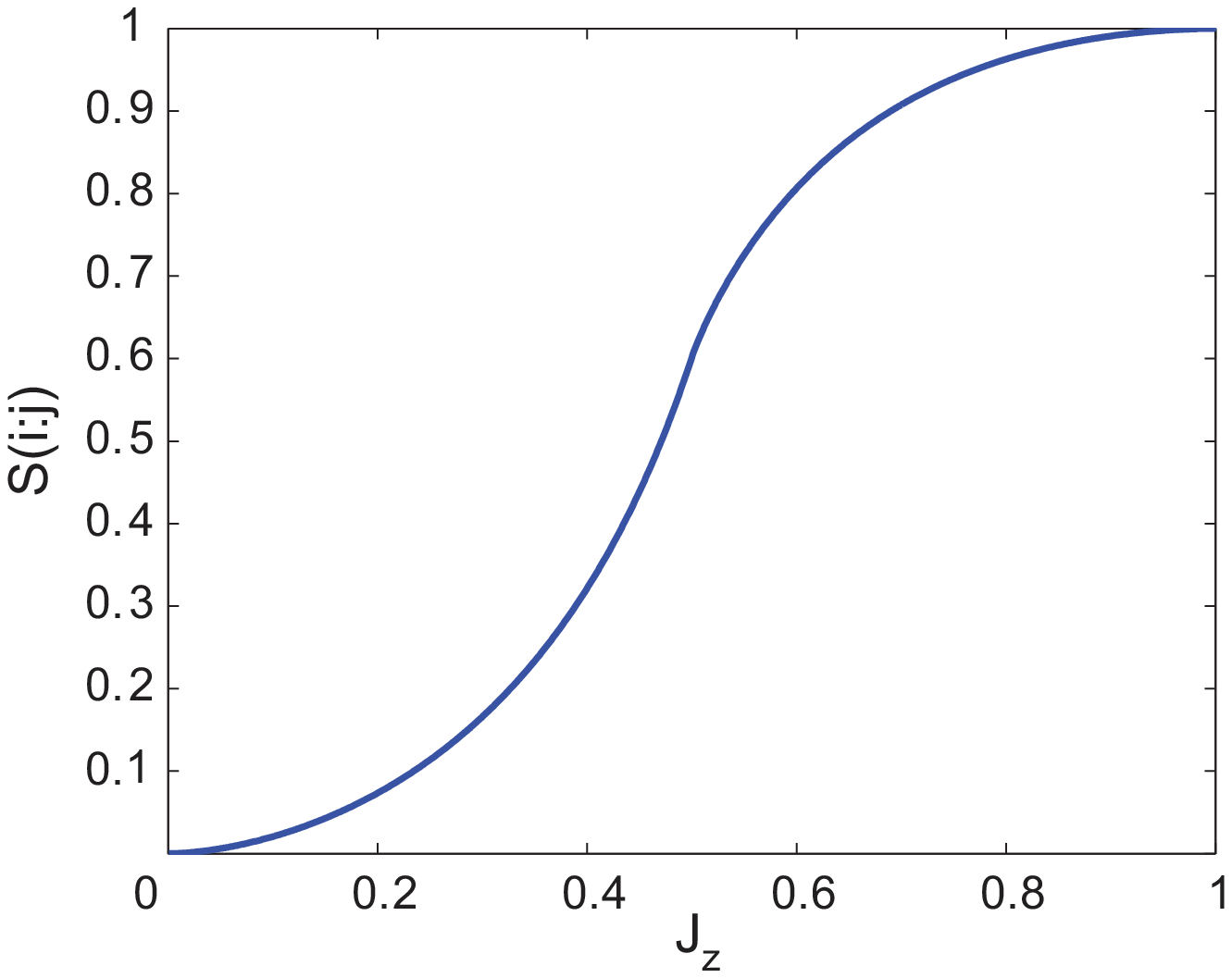}
\caption{\label{fig:epsart}(color online). Mutual information
between two connected sites. The mutual information increases
monotonically with the increasing of $J_z$. However, there exist a
certain value $J_z^m$. If $J<J_z^m$, the mutual information is a
concave function while if $J>J_z^m$, the mutual information is a
convex function. Thus, the first order derivative of the mutual
information with respect to $J_z$ has a peak at the $J_z^m$. $J_z$
is in unit of $J_x+J_y+J_z$.}
\end{figure}
\begin{figure}
\includegraphics[height=11cm,width=13cm]{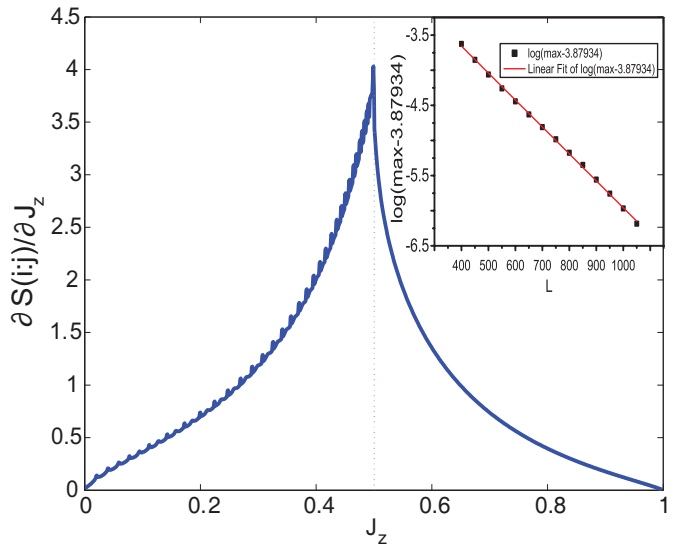}
\caption{\label{fig:epsart}(color online). The first order
derivative of mutual information between two connected sites for the
system-size $L=100$. The derivative of the mutual information has a
maximum exactly at the critical point $J_z=0.5$. The small peaks in
the gapless region is due to that while calculating the
$\langle\sigma_{\textbf{r},1}^{z}\sigma_{\textbf{r},2}^{z}\rangle$
certain $\varepsilon_{\textbf{q}}$ and $E_{\textbf{q}}$ are both
infinitesimal and some systemical error are included. The peaks get
smaller when $L$ increases. The subgraph shows that when the
system-size tends to infinity, the maximum of the derivative of the
mutual information tends to a constant as
$\log_2\Big(\big(\frac{\partial S(i:j)}{\partial
J_z}\big)_{max}-3.87934\Big)=-0.00384L-2.12174$. $J_z$ is in unit of
$J_x+J_y+J_z$.}
\end{figure}
\begin{figure}
\includegraphics[height=7cm,width=\linewidth]{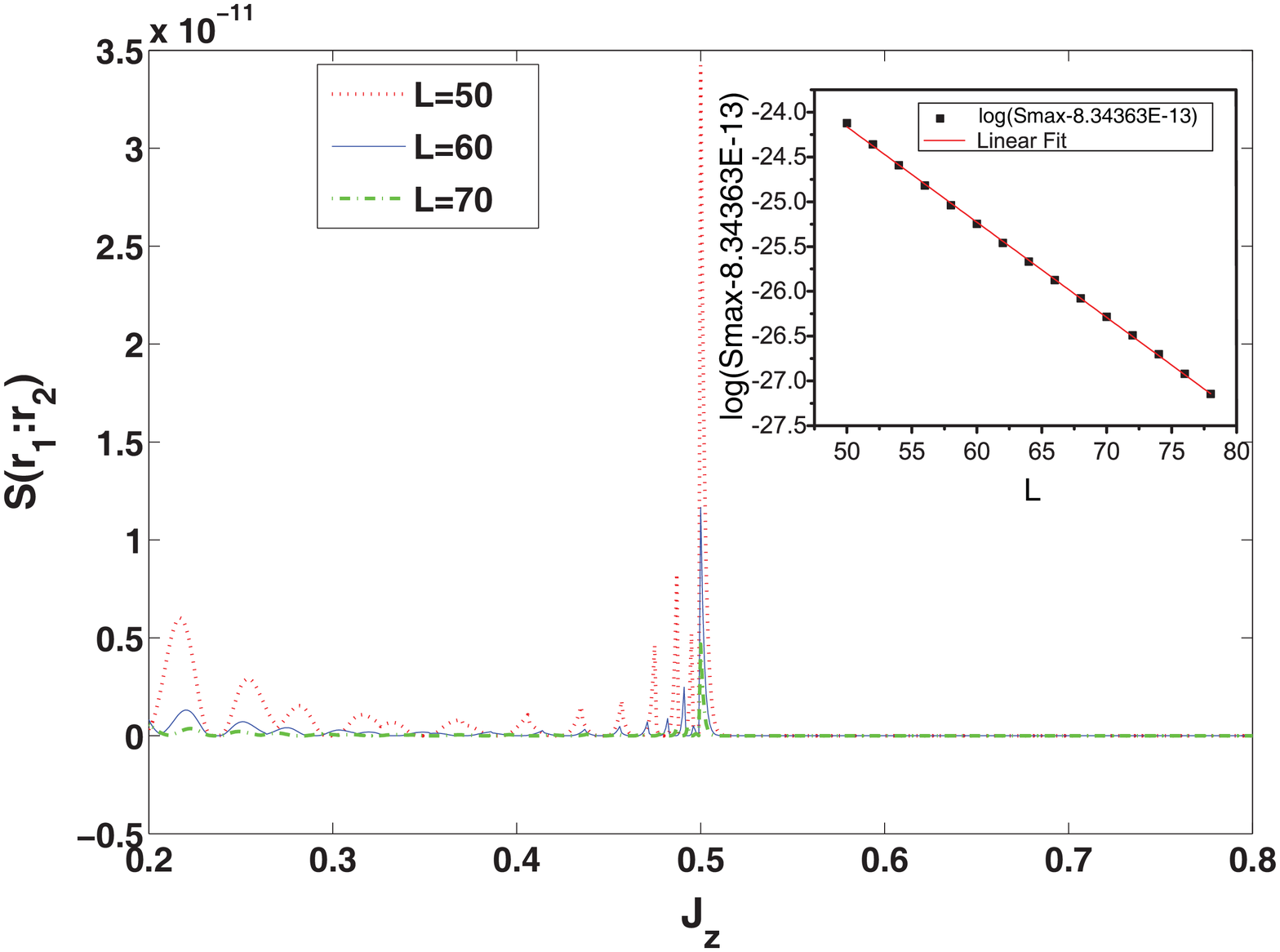}
\caption{\label{fig:epsart}(color online). The mutual information
between two $z$ linked bonds with longest distance located on the
torus surface for a given system-size. The small peaks in the
gapless phase is caused by the $0/0$ error as before. The
significant peak arises exactly at the critical point. The subgraph
shows that when the system-size tends to infinity, the peak value of
the mutual information trends to a constant as
$\log_2\big(S(\textbf{r}_1:\textbf{r}_2)_{max}-8.34363\times10^{-13}\big)=-0.10637
L-18.8454 $. Here $J_z$ is in unit of $J_x+J_y+J_z$.}
\end{figure}

In order to show our results more clearly, we use the numerical
method to diagonalize the Hamiltonian exactly. With the periodical
boundary conditions, we diagonalize an eight lattices system which
is the smallest subsystem on the surface of a torus, please see
figure $3$, and calculate all the $16$ two-site correlation
functions. Here we omit the subindex corresponding to the sites'
position because the value is invariant due to the translational
invariance of this system. The explicit form of the Hamiltonian
reads
\begin{eqnarray}
H_8&=&J_x(\sigma^x_5\sigma^x_3+\sigma^x_6\sigma^x_4+\sigma^x_7\sigma^x_1+\sigma^x_8\sigma^x_2)\nonumber\\
&+&J_y(\sigma^y_3\sigma^y_6+\sigma^y_5\sigma^y_4+\sigma^y_8\sigma^y_1+\sigma^y_7\sigma^y_2)\nonumber\\
&+&J_z(\sigma^x_3\sigma^x_7+\sigma^x_4\sigma^x_8+\sigma^x_5\sigma^x_1+\sigma^x_6\sigma^x_2).
\label{new}
\end{eqnarray}
By using the periodical boundary conditions, the system-size of the
system (\ref{new}) can be extend to infinity. The main properties of
the system are kept since all possible interactions are considered.
The result is shown in figure $4$. We see that the correlation
functions along the $z$-direction is non-zero.

The explicit form of reduced density matrix of two sites with
nearest neighbor is
\begin{widetext}
\begin{equation}
\begin{array}{cccc}
\rho_{\textbf{r},1;\textbf{r},2}=\frac{1}{4}\left(\begin{array}{cccccccc}&1+\langle\sigma^z_{\textbf{r}_1,1}\sigma^z_{\textbf{r}_1,2}\rangle&0&0&0\\
\\
&0 &1-\langle\sigma^z_{\textbf{r}_1,1}\sigma^z_{\textbf{r}_1,2}\rangle &0 &0\\
\\
&0 &0 &1-\langle\sigma^z_{\textbf{r}_1,1}\sigma^z_{\textbf{r}_1,2}\rangle &0\\
\\
&0 &0 &0 &1+\langle\sigma^z_{\textbf{r}_1,1}\sigma^z_{\textbf{r}_1,2}\rangle\\
\end{array}
\right)
\end{array}.
\end{equation}
\end{widetext}
The eigenvalues of the reduced density matrix are
$\lambda_1=\lambda_2=(1-\langle\sigma^z\sigma^z\rangle)/4$ and
$\lambda_3=\lambda_4=(1+\langle\sigma^z\sigma^z\rangle)/4$. Every
eigenvalue corresponds the two-fold degenerate eigenstates. From the
two sites reduced density matrix we can derive the reduced density
matrix for one site as $I/2$, where $I$ is identity matrix. The
mutual information of the two sites $i$ and $j$ is
\begin{eqnarray}
S(i:j)=2-2H\Big(\frac{1-\langle\sigma^z\sigma^z\rangle}{4}\Big)-2H\Big(\frac{1+\langle\sigma^z\sigma^z\rangle}{4}\Big),
\end{eqnarray}
where $H(x)=-x\log_2(x)$.

The the two-site mutual information along the line
$J_x=J_y=(1-J_z)/2$ is shown in figure 5. We see that the mutual
information increases monotonically with the increasing of $J_z$.
However, there exist a certain value $J_z^m$. If $J<J_z^m$, the
mutual information is a concave function while if $J>J_z^m$, the
mutual information is a convex function. Thus, the first order
derivative of the mutual information with respect to $J_z$ has a
peak at the $J_z^m$. The derivative of the mutual information is
shown in figure 6. From figure 6, we see that the first order
derivative of the mutual information arrives at the maximum value at
the point $J_z^m=0.5$, which exactly corresponds the critical point
$J_z=0.5$. We also find that the value of $J_z^m$ is fixed when the
system-size changes. The maximum value is a constant when the system
size tends to infinity, as shown in figure 7. This is different from
the Ising model where the second order derivative of entanglement
entropy diverges at the critical point in the the thermodynamic
limit\cite{cao}, although their density matrixes are same in
structure.

From the quantum information perspective, the entanglement measured
by concurrence between two sites is zero since the density has only
diagonal elements, while the entanglement between one site and all
the rest sites is maximum entanglement, i.e.,
$c=\sqrt{\frac{d}{d-1}(1-Tr\rho_i^2)}=1$, where $c$ denotes the
concurrence, $\rho_i$ is the reduced density matrix of the particle
in site $i$ and $d$ is the dimension of
$\rho_i$\cite{asfsm,wdcirac}.

\section{\label{sec:level1}M\lowercase{utual information between two bonds with longest distance}  }

In this section, we investigate the mutual information between two
$z$ linked bonds. Firstly, we need to calculate the density matrix.
For arbitrary two $z$ linked bonds at $\textbf{r}_1$ and
$\textbf{r}_2$ the density matrix is
\begin{eqnarray}
\rho_{\textbf{r}_1,\textbf{r}_2}&=&\frac{1}{16}\sum_{\alpha,\beta=0,3}\sigma^{\alpha}_{\textbf{r}_1,1}
\sigma^{\alpha}_{\textbf{r}_1,2}\sigma^{\beta}_{\textbf{r}_2,1}\sigma^{\beta}_{\textbf{r}_2,2}
\nonumber \\
&& \quad \times
\langle\sigma^{\alpha}_{\textbf{r}_1,1}\sigma^{\alpha}_{\textbf{r}_1,2}\sigma^{\beta}_{\textbf{r}_2,1}\sigma^{\beta}_{\textbf{r}_2,2}\rangle.
\end{eqnarray}
The eigenvalues of this density matrix are
$(1-4\langle\sigma^z\sigma^z\sigma^z\sigma^z\rangle)/16$,
$(1-2\langle\sigma^z\sigma^z\rangle+\langle\sigma^z\sigma^z\sigma^z\sigma^z\rangle)/16$,
$(1+2\langle\sigma^z\sigma^z\rangle+\langle\sigma^z\sigma^z\sigma^z\sigma^z\rangle)/16$.
These eigenvalues correspond the $8$ fold, $4$ fold and $4$ fold
degenerate eigenstates, respectively. Then we obtain the mutual
information between two $z$ linked bonds in $\textbf{r}_1$ and
$\textbf{r}_2$ as
\begin{eqnarray}
S(\textbf{r}_1:\textbf{r}_2)&=&4H\left(\frac{1+\langle\sigma^z\sigma^z\rangle}{4}\right)
+4H\left(\frac{1-\langle\sigma^z\sigma^z\rangle}{4}\right)\nonumber\\
&-&8H\left(\frac{1-\langle\sigma^z\sigma^z\sigma^z\sigma^z\rangle}{16}\right)\nonumber\\
&-&4H\left(\frac{1-2\langle\sigma^z\sigma^z\rangle+\langle\sigma^z\sigma^z\sigma^z\sigma^z\rangle}{16}\right)\nonumber\\
&-&4H\left(\frac{1+2\langle\sigma^z\sigma^z\rangle+\langle\sigma^z\sigma^z\sigma^z\sigma^z\rangle}{16}\right).
\end{eqnarray}

In order to reveal the long range correlation in the system, we
study the bonds in the direction that contains the largest distance
in the torus, which is marked by a line of elliptic circles in
figure 1. Without losing generality, we choose the positions of the
bonds as $\textbf{r}_1=(0,0)$ and $\textbf{r}_2=(0.5L,0.5L)$ in the
$(\textbf{n}_1,\textbf{n}_2)$ coordinate system so that the two
bonds are longest separated. The mutual information between two $z$
linked bonds in this direction is shown in figure 7. We find that
the mutual information has a maximum at the critical point. This
property is valid for different system-sizes. The peak values tend
to a constant when the system-size $L$ trends to infinity.

\section{\label{sec:level1}C\lowercase{onclusions }}

In this paper, based on the exact ground state of Kitaev honeycomb
model we have obtained both the reduced density matrices of two
nearest neighbor sites and between two $z$ linked bonds with longest
distance. We show that these density matrices have only diagonal
elements, so that there is no entanglement between two local sites
or two local bonds, but the nonlocal entanglement between one site
and the rest of the whole system is maximum. From quantum
information theory, the ground state which is a multipartite state
seems more like  Greenberger-Horne-Zeilinger (GHZ) states other than
W-type states. We have calculated the mutual information between two
nearest neighbor sites and the mutual information between two $z$
linked bonds with the longest distance in the torus topology. The
first order derivative of the former mutual information and the
latter mutual information itself have peak at the point where the
ground state transits from the gapless phase into a gapped phase.
The above singular behavior serves as an exact and easily obtaining
detector of the topological phase transition in the Kitaev honeycomb
model. The so called \emph{localizable entanglement} is related with
the string order parameters and the hidden topological long range
order in one dimension spin chains\cite{mpfvmamdjid,hfzdwvv}. More
over, the topological phase and topological phase transition have
their roots in the hidden topological long range order and the
string order parameters. Therefore, the investigation of the
relation among the localizable entanglement, topological phase
transition and other related quantities in the exactly solvable two
dimensional Kitaev model may greatly enhance our understanding of
the topological phase. The research in this direction should be
further explored to extensively study the topological phase and
topological order by means of quantum information.

\section{\label{sec:level1}A\lowercase{cknowledgement}}

This work is supported by grants of National Natural Science
Foundation of China and ``973'' program of Ministry of Science and
Technology (MOST), China.

\end{document}